\documentclass{PoS}

\usepackage{slashbox}
\usepackage{bbm}
\pdfoutput=1

\newcommand{\val}{\mathrm{v}}
\newcommand{\sea}{\mathrm{s}}
\newcommand{\be}{\begin{equation}}
\newcommand{\ee}{\end{equation}}
\newcommand{\tr}{\mathrm{Tr} \,}

\newcommand{\dd}[1]{\mathrm{d} #1 \,}

\newcommand{\id}{\ensuremath{\mathbbm{1}}}
\newcommand{\eff}{\mathrm{eff}}
\newcommand{\g}{\gamma}

\newcommand{\vev}[1]{\ensuremath{\left\langle #1 \right\rangle}}

\title{$N_f=2$ chiral dynamics in the mixed chiral regime}

\ShortTitle{$N_f=2$ chiral dynamics in the mixed chiral regime}

\author{\speaker{Gr\'egory Vulvert}, Pilar Hern\'andez \\ 
        Instituto de F\'isica Corpuscular, CSIC -- Universitat de Val\`encia \\ 
        Apartado de Correos 22085, E -- 46071 Valencia, Spain.\\
        E-mail: \email{gregory.vulvert@ific.uv.es}, \email{pilar.hernandez@ific.uv.es}}

\author{Silvia Necco\\
	Albert Einstein Center for Fundamental Physics, \\
	Institute for Theoretical Physics \\
	Sidlerstrasse 5, CH -- 3012 Bern, Switzerland. \\
    E-mail: \email{necco@itp.unibe.ch}}

\author{Carlos Pena\\
        Dpto. de F\'isica T\'eorica and Instituto de F\'isica T\'eorica UAM/CSIC\\
 		Universidad Aut\'onoma de Madrid, Cantoblanco E -- 28049 Madrid, Spain.\\       
        E-mail: \email{carlos.pena@uam.es}}

\abstract{We present a study of the pseudoscalar propagator in the mixed chiral regime with valence quark masses in the $\epsilon-$regime and sea quark masses in the $p-$regime. We first show the NNLO prediction of this observable in the chiral expansion. In sectors of fixed topology, the correlator has a pole in $1/m_\val^2$ that can be matched to the topological zero-mode contributions to the correlator. We compute the residue of this pole  in  a $N_f=2$ mixed-action simulation and compare the results with the prediction in Chiral Perturbation Theory (ChPT).}

\FullConference{The 30th International Symposium on Lattice Field Theory\\
                 June 24 -- 29,  2012\\
                 Cairns, Australia}

\begin{document}

\section{Introduction}

Lattice QCD allows to extract physical results from unphysical conditions such as finite volume or unphysical quark masses. In particular the up and down quark masses are typically heavier in lattice simulations that in nature. Nevertheless chiral extrapolations allow to extract physical results. A more extreme  generalization  is the  partially-quenched setup where quark masses are different in the valence and sea sectors.

It is also possible to use different discretizations in the valence and sea sectors: these are  the mixed-action formulations~\cite{Bar2003, Bar2004}.  One such setup is to consider overlap valence quarks on improved-Wilson sea, so that an exact valence quark chiral symmetry can be preserved at a moderate cost. This symmetry can be very useful to deal with renormalization of four-fermion operators or to access other kinematical regimes such as the $\epsilon-$regime.  

In this work we present the result for the pseudoscalar propagator  at NNLO in the mixed-regime of ChPT where some quarks are in the $\epsilon-$regime and some in the $p-$regime~\cite{Bernardoni2007, Bernardoni2008}. The same computation was done with all quarks in the $\epsilon-$regime in~\cite{Giusti2004}. 
When valence quarks are in the $\epsilon$-regime, there are exact poles in $1/m_\val^2$ when the averages are considered in fixed topological sectors.  The residue of these poles can on the one hand be computed in ChPT  in terms of low-energy couplings, and on the other computed on the lattice from  the wave-functions of the topological zero-modes. The latter being potentially more efficient than methods requiring the computation of the full correlation function~\cite{Giusti2004}. 

We have computed these observables in an exploratory mixed action simulation with valence overlap fermions on a sea of $N_f$ non-perturbatively $\mathcal{O}(a)-$improved Wilson fermions (generated within the CLS common effort~\footnote{https://twiki.cern.ch/twiki/bin/view/CLS/WebHome.}). We comment on the comparison of the numerical results and the expectation from ChPT.

\section{Topological zero-modes wave functions}

\begin{figure}
\begin{center}
\includegraphics[scale=0.3]{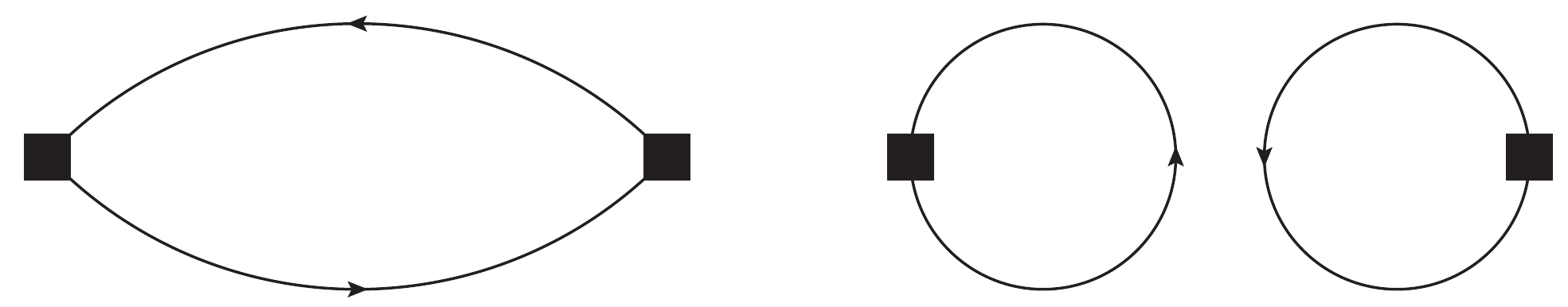}
\end{center}
\caption{Illustration of the two types of quark lines contractions. On the left, the connected contraction, corresponding to the amplitude $\mathcal{A}$. On the right the disconnected one, corresponding to the amplitude $\tilde{\mathcal{A}}$.}
\label{fig:quarklines}
\end{figure}

Writing the spectral decomposition of the quark propagator 
\be 
D^{-1}_{xy} = \sum_{i, \atop \mathrm{zero-modes}} \frac{v_i(x)v_i^\dagger(y)}{m V} + \sum_{i, \atop \mathrm{non \; zero-modes}} \frac{v_i(x)v_i^\dagger(y)}{\left(\lambda_i+ m \right) V},
\ee 
it can be shown that a two-point function, computed at fixed topology $\nu$ (indicated by $\langle \cdots \rangle_\nu$) contains a pole in $m_\val^2$ due to exact zero-modes. Its residue is given by:
\be 
\lim_{m_\val \to 0} \left( m_\val V \right)^2 C^{ab}_\nu \left( x \right) = \mathrm{Tr} \, \left[ T^a T^b \right] \mathcal{A}_\nu(x) + \mathrm{Tr} \, [ T^a ] \mathrm{Tr} \; [ T^b] \tilde{\mathcal{A}}_\nu(x) \label{eq:PPspectralrep}
\ee
where 
\be
\mathcal{A}_\nu (x-y)  =  \left\langle \sum_{i,j} v_j^\dagger(x) v_i(x) v_i^\dagger(y) v_j(y) \right\rangle_\nu, \,
\tilde{\mathcal{A}}_\nu (x-y)  =  -\left\langle \sum_{i} v_i^\dagger(x) v_i(x) \sum_j v_j^\dagger(y) v_j(y) \right\rangle_\nu. \label{eq:Adisc}
\ee
The sums run over the set of $|\nu|$ zero modes $v_i$ of the Dirac operator, $D v_i =0$, which have definite chirality, and are assumed to be normalized so that
$\displaystyle{\int \dd{x}~  v_i^\dagger(x) v_i(x) = V}$.
The left side of eq.~(\ref{eq:Adisc}) corresponds to a ``connected'' contraction of the quark lines whereas the right part corresponds to a ``disconnected'' one. The figure~(\ref{fig:quarklines}) illustrates these features.
It is more efficient to compute the residues themselves than the full all-to-all propagator, so it makes sense to use these observables in the matching to ChPT. 

\section{Pseudoscalar correlator in ChPT in the mixed regime}

We consider QCD with $N_\val$ valence quarks of mass $m_\val$ and $N_\sea$ sea quarks of mass $m_\sea$.
The two-point function of the pseudoscalar density that we want to compute is defined by:

\be\label{eq:PPdef2}
\tr [T^a T^b] \, C^{ab}_\nu(t) = \int \dd{\vec{x}}~\vev{P^a(x) P^b(0)}_\nu, \quad P^a(x) \equiv  \bar{\psi}_\val T^a i \g_5 \psi_\val,
\ee
where $T^a,~a=1, \ldots,~N_\val^2-1$ is a traceless generator of $SU(N_\val)$ and $T^0 = \id_\val$.
ChPT can predict the behaviour of the pseudoscalar correlator in finite volume $V=L^3 T$ and as a function of the quark masses and of a finite number of low-energy couplings.

In the mixed regime we consider, the valence quarks are in the $\epsilon-$regime and the sea quarks  in the $p-$regime:
\be 
m_\val \Sigma V \lesssim 1 \; \textrm{ and } \; m_\sea \Sigma V \gg 1.
\ee
The power counting is thus the following:
\be 
p^2 \sim L^{-2} \sim m_\sea \sim \mathcal{O}(\epsilon^2) \quad \mathrm{and} \quad m_\val \sim \mathcal{O}(\epsilon^4) .
\ee
In the $\epsilon-$regime, the only coupling appearing at NNLO is $F$. However, in the mixed regime, the higher order couplings $L_4, \ldots, L_8$ contribute because of the heavier sea quarks in the $p-$regime. The sea quarks  therefore behave as decoupling particles: we expect that the correlator~(\ref{eq:PPdef2}) has the same behavior as the one in the quenched theory with $\epsilon-$regime quarks, but with effective low-energy couplings which are well-defined functions of those of the $N_f=N_\sea$ theory. Indeed this expectaction is satisfied.
We refer to~\cite{Bernardoni2007, Bernardoni2008} for more details  on the Feynman rules in the mixed-regime. 

More concretely, we consider  the volume average of the euclidean time derivative of the residues of the correlators:

\begin{eqnarray*}
\lim_{m_\val \to 0}  \left(m_\val V \right)^2 \frac{\mathrm{d}}{\mathrm{d}t} \int \mathrm{d}^3 \vec{x}~ C^{aa}_\nu (x) & = & \frac{1}{2} A_\nu^\prime(t), \\
\lim_{m_\val \to 0}  \left(m_\val V \right)^2 \frac{\mathrm{d}}{\mathrm{d}t} \int \mathrm{d}^3 \vec{x}~ C^{00}_\nu (x) & = & N_\val A_\nu^\prime(t) + N_\val^2 \tilde{A}_\nu^\prime (t).
\end{eqnarray*}

The spectral representation of~(\ref{eq:PPspectralrep}) and the definitions $\displaystyle{\mathcal{A}_\nu(t) = \int \dd{\vec{x}} \mathcal{A}_\nu(x)}$ and $\displaystyle{\tilde{\mathcal{A}}_\nu(t) = \int \dd{\vec{x}} \tilde{\mathcal{A}}_\nu(x)}$ implie that a large $t$: $\displaystyle{\mathcal{A}_\nu^\prime(t) = A_\nu^\prime(t)~\textrm{ and }~\tilde{\mathcal{A}}_\nu^\prime(t) = \tilde{A}_\nu^\prime(t)}$.

We have computed these quantities at NNLO for any $N_\val$ and $N_\sea$, but we will focus here on the partially-quenched case ($N_\val=0$), which is the relevant one for the numerical study of this work.

\section{ChPT predictions in the partially-quenched theory}

We use the replica method~\cite{Damgaard2000} : the perturbative modes give contributions that are finite in the limit $N_\val \to 0$, while 
the contribution of the non-perturbative modes  involve zero-mode integrals on the $U(N_\val |N_\val)$ supergroup. We checked explicitely that the results obtained are consistent with the ones obtained directly from the supersymmetric method~\cite{Damgaard2002}. 

We obtain the following structure for the correlator: 
\begin{eqnarray} 
C^{aa}_q(x) &  = &  C_q + \alpha_q^{(1)} \bar{G}(x,0) + \alpha_q^{(2)} \bar{F}(x) + \beta_q^{(1)} \left[\bar{G}(x,0)  \right]^2 + \beta_q^{(2)} \left[\bar{F} \left(x \right) \right]^2 + \beta_q^{(4)} \bar{G}(x,0) \bar{F}\left(x \right) \nonumber \\
              &   & + {} \; \int \mathrm{d}^4 y \, \left[ \gamma_q^{(1)} \bar{G}(x-y,0) \bar{G}(y,0)  +  \gamma_q^{(2)} \bar{G}\left(x-y,0 \right) \bar{F}\left(y \right) + \gamma_q^{(3)} \bar{F}\left(x-y\right) \bar{F}\left(y \right) \right] \\
              &  & + {} \; \beta_q^{(3)} \left[\bar{G}\left(x, M_{\rm{ss}}^2 /2\right)\right]^2   + \varepsilon_q \delta^{(4)}(x), \nonumber
\end{eqnarray}
with the following definition of the propagators:
\begin{eqnarray}
G \left(x, M^2\right) & = & \frac{1}{V} \sum_{n \in \mathbb{Z}^4} \frac{\mathrm{e}^{i p \cdot x}}{p^2 + M^2}, \quad \bar{G} \left(x \right)  =  \frac{1}{V} \sum_{n \in \mathbb{Z}^4}  \left( 1 - \delta^{(4)}_{n,0} \right) \frac{\mathrm{e}^{i p \cdot x}}{p^2}, \\
\bar{F} \left(x\right) & = & \frac{1}{V} \sum_{n \in \mathbb{Z}^4} \left( 1 - \delta^{(4)}_{n,0} \right) \frac{\mathrm{e}^{i p \cdot x}}{p^4}, \quad p=2 \pi \left( \frac{n_0}{T}, \frac{\vec{n}}{L}\right).
\end{eqnarray}

Except for the first term of the third line, the result has the same functional form as the fully-quenched result in the $\epsilon-$regime~\cite{Giusti2004}. However the coefficients are different: they are now functions of the low-energy couplings of the $N_f = N_\sea$ theory and of the quark masses, as expected. In particular, the singlet couplings $m_0^2$ and $\alpha$ appearing explicitly in the quenched case do not show up in this case since the singlet is integrated out in the partially-quenched theory. However, both results match provided the following identifications are used:

\vspace*{-5mm}
\begin{eqnarray}
F_{\rm{eff}}^2 & = & F^2 \left\{ 1 - \frac{N_\sea}{F^2} \left[ G\left(0, M_{\rm{ss}}^2 /2 \right) - 8 L_4 M_{\rm{ss}}^2 \right]\right\} \\
\frac{\alpha_{\rm{eff}}}{2 N_c} & = & \frac{1}{N_\sea} \left\{ 1 - \frac{1}{F^2} \left[ N_\sea G\left(0, M_{\rm{ss}}^2 /2 \right) - 8 L_5 M_{\rm{ss}}^2 \right]\right\} \label{eq:alphaeff} \\
\frac{m_{0, \, \rm{eff}}^2}{2 N_c} & = &   \frac{M_{\rm{ss}}^2}{N_\sea} \left\{ 1 - \frac{1}{F^2} \left[ \frac{N_\sea^2 - 1}{N_\sea}G\left(0, M_{\rm{ss}}^2 \right) - 16  M_{\rm{ss}}^2 \left( N_\sea L_6  + N_\sea L_7 + L_8 \right) \right] \right\} \label{eq:m0eff}
\end{eqnarray}

In~\cite{Giusti2004}, a large$-N_c$ expansion was combined with the chiral expansion so that $1/N_c \sim \mathcal{O}(\epsilon^2)$, and the scaling of the singlet couplings was chosen to be $m_0^2/ N_c \sim M_{\rm{ss}}^2 \mathcal{O}(\epsilon^4)$ and $\alpha/ N_c \sim \mathcal{O}(\epsilon^2)$. We do not get the same power counting since eqs.~(\ref{eq:alphaeff}--\ref{eq:m0eff}) imply that $m_{0, \, \rm{eff}}^2/ N_c \sim \mathcal{O}(\epsilon^2)$ and $\alpha_\eff/ N_c \sim \mathcal{O}(1)$. To achieve the same power counting we would need to assume $1/N_\sea  \sim \mathcal{O}(\epsilon^2)$. Under this assumption, the present results precisely match with the fully-quenched ones of~\cite{Giusti2004}, as expected. 

A useful quantity to match lattice QCD results is $D_\nu$ so that 
\be
\frac{1}{L^2} A_\nu^\prime(t) \equiv D_\nu z  + C_\nu z^3 + \mathcal{O}(z^5), \quad z=\tau - \frac{1}{2} \, \textrm{ and } \, \tau = \frac{t}{T},
\label{eq:tdep}
\ee
and similarly for $\tilde{D}_\nu$. The study of the quenched case~\cite{Giusti2004} showed that $D_\nu$ can be robustly extracted from a fit of this form to the temporal dependance of the time derivative of the correlators. The ChPT prediction for $D_\nu$ is:
\begin{eqnarray}
D_\nu & = & \frac{2 |\nu|}{\left( F_{\rm{eff}} L \right)^2} \left\{ |\nu| + \frac{\alpha_{\rm{eff}}}{2 N_c} + \frac{\rho}{\left( F_{\rm{eff}} L \right)^2} \left[ -\beta_1 \rho^{-3/2} + \left( \frac{5}{N_\sea ^2} + \frac{8 |\nu| }{N_\sea} + 3 + 2\nu^2 - 2\langle \nu^2 \rangle_{\rm{eff}} \right) \zeta_2 \right. \right.   \nonumber \\
      &   & + {} \; \left. \left.   \left( \frac{1}{N_\sea ^2} + \frac{2 |\nu|}{N_\sea}+\frac{1}{2}\right) \gamma_1 + 2 \frac{M_{\rm{ss}}^2}{N_s}T^2\left( \frac{1}{N_\sea} + |\nu| \right)\left(\gamma_2 -4 \zeta_3\right) + \left( \frac{M_{\rm{ss}}^2}{N_\sea} \right)^2 T^4 \left( \gamma_3 + 4 \zeta_4 \right)  \right. \right. \nonumber \\
      &   & - {} \; \left. \left. \frac{N_\sea}{2} |\nu| \gamma_4 \left(M_{\rm{ss}}^2/2 \right) \right] \right\} \label{eq:Dnudep}
\end{eqnarray}
with  $\displaystyle{\langle \nu^2 \rangle_{\rm{eff}} = V \frac{m_{0, {\rm{eff}}}^2 F^2_{\rm{eff}} }{4 N_c}}$ (as in the Witten-Veneziano relation).
We define here $\rho = T/L$; and the numerical values for $\beta_1$, $\zeta_i$ and $\gamma_i$ for $i=1, \ldots, 4$ are given in table~(\ref{table:coeffgeo}).

\begin{table}[h!]
   \hspace*{-4mm}
   \begin{minipage}[b]{0.5\linewidth}
      \centering 
      \renewcommand{\arraystretch}{0.9}
	\begin{tabular}{|c|c|c|}
	\hline
        \backslashbox{\hspace*{2mm}Coeff}{\hspace*{-2.2mm}\vspace*{-2.6mm}$\rho$}      &  $1$ & $2$ \\
        \hline\hline
	$\beta_1$ & $0.140461$ & $0.083601$ \\ \hline  
	$\zeta_1$ & \multicolumn{2}{c|}{$1$}   \\ 
    $\zeta_2$ & \multicolumn{2}{c|}{$-1/24$} \\ 
	$\zeta_3$ & \multicolumn{2}{c|}{$7/5760$}  \\ 
	$\zeta_4$ & \multicolumn{2}{c|}{$-31/967680$} \\ \hline 
	$\gamma_1$ & $-0.057128$ & $-0.083291$      \\ 
    $\gamma_2$ & $-0.001951$ & $-0.002951$         \\ 
	$\gamma_3$ & $-0.000066$ & $-0.000101$      \\ \cline{2-3}
	$\gamma_4$ & \multicolumn{2}{c|}{pion sea mass dependent}       \\ \hline
	\end{tabular}
\caption{\label{table:coeffgeo}}{Geometry dependent coefficients.}
   \end{minipage}
   \hspace*{-4mm}
   \begin{minipage}[b]{0.5\linewidth}  
    \centering
	\renewcommand{\arraystretch}{0.9}
    \begin{tabular}{|c|c|cccc|}
	\hline
	Lattice & Total \# & \multicolumn{4}{c|}{\# conf. by topological sector} \\ \cline{3-6}
                &    conf.           & $\nu=1$ & $\nu=2$ & $\nu=3$ & $\nu=4$ \\
        \hline\hline
	D4 & $117$ & $43$ & $18$ & $31$ & $25$ \\ \hline 
	D5 & $129$ & $40$ & $43$ & $23$ & $23$ \\ \hline
	D6 & $153$ & $85$ & $43$ & $25$ & -- \\ \hline	
	\end{tabular}
	\vspace{10mm}
	\caption{\label{table:stat}}{Data sample.}
    \end{minipage}
\end{table}

\section{Numerical results}

We have used $N_\sea=2$ CLS $\mathcal{O}(a)-$ improved gauge configurations on which we built overlap valence fermions in a  volume of ${24^3 \times 48}$ at $\beta=5.3$, which corresponds to a value of the lattice spacing $a=0.0649(10) \; \textrm{ fm}$~\cite{Marinkovic2011}. Equivalentely, these values correspond to a lattice size $L = 1.56\; \textrm{ fm}$ for $\rho = 2$.
We use three lattices D4, D5 and D6 corresponding to $\kappa = 0.13620,~ 0.13625$ and $0.136635$ respectively.

\begin{table}[h]
	\centering
	\renewcommand{\arraystretch}{0.9}
	\begin{tabular}{|c|c|cccc|}
	\hline
        Lattice & Fitrange & $\nu=1$ & $\nu=2$ & $\nu=3$ & $\nu=4$ \\
        \hline\hline
D4 & $24-33$ & $3.0(0.6)$ & $8.9(0.6)$ & $11.5(0.8)$ & $16.6(1.6)$ \\ 
   & $24-35$ & $3.0(0.5)$ & $8.6(0.5)$ & $11.7(0.7)$ & $17.5(1.3)$ \\ 
   & $24-37$ & $2.9(0.5)$ & $8.6(0.5)$ & $12.0(0.7)$ & $18.5(1.2)$ \\  
   & $24-39$ & $2.9(0.4)$ & $8.9(0.5)$ & $12.3(0.7)$ & $19.5(1.1)$ \\  \hline

D5 & $24-33$ & $2.1(0.9)$ & $6.5(0.9)$ & $11.9(1.3)$ & $17.8(1.4)$ \\ 
   & $24-35$ & $2.7(0.7)$ & $7.1(0.8)$ & $13.5(1.0)$ & $19.0(1.0)$ \\ 
   & $24-37$ & $3.1(0.6)$ & $7.7(0.7)$ & $14.6(0.9)$ & $20.0(0.9)$ \\  
   & $24-39$ & $3.3(0.6)$ & $8.3(0.6)$ & $15.3(0.9)$ & $21.2(0.9)$ \\  \hline

D6 & $24-33$ & $4.4(0.6)$ & $9.8(1.0)$  & $16.3(1.3)$ & -- \\ 
   & $24-35$ & $4.5(0.5)$ & $10.0(0.9)$  & $16.7(1.1)$ & -- \\ 
   & $24-37$ & $4.6(0.5)$ & $10.5(0.8)$ & $17.2(1.0)$ & -- \\  
   & $24-39$ & $4.7(0.4)$ & $11.1(0.7)$ & $17.8(1.0)$ & -- \\  \hline
	\end{tabular}	
\caption{\label{table:Dres}}{$D_\nu$ extracted from the linear fit of $A^\prime(t)/L^2$ for several time intervals.}
\end{table}

The reader can refer to~\cite{Bernardoni2011} were topological and spectral observables have been studied on the same set on configurations. Thus these are already classified by topological charge and we could therefore compute the zero-mode saturated correlators at fixed topological charge. Table~(\ref{table:stat}) summarizes the data sample.

\begin{figure}[h!]
   \hspace*{-6mm}
   \begin{minipage}[b]{0.5\linewidth}
      \centering \includegraphics[scale=0.8]{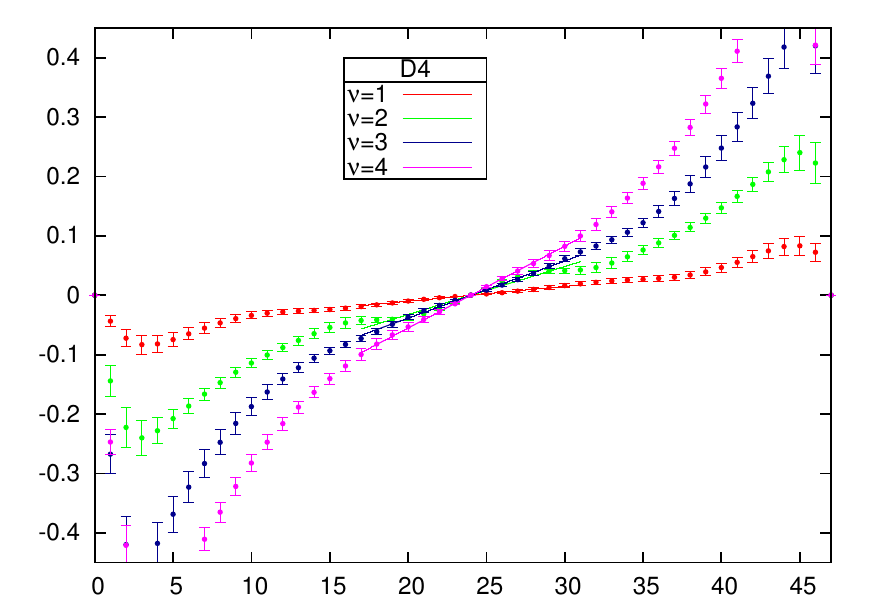}
   \end{minipage}
   \hspace*{3mm}
   \begin{minipage}[b]{0.5\linewidth}   
      \centering \includegraphics[scale=0.8]{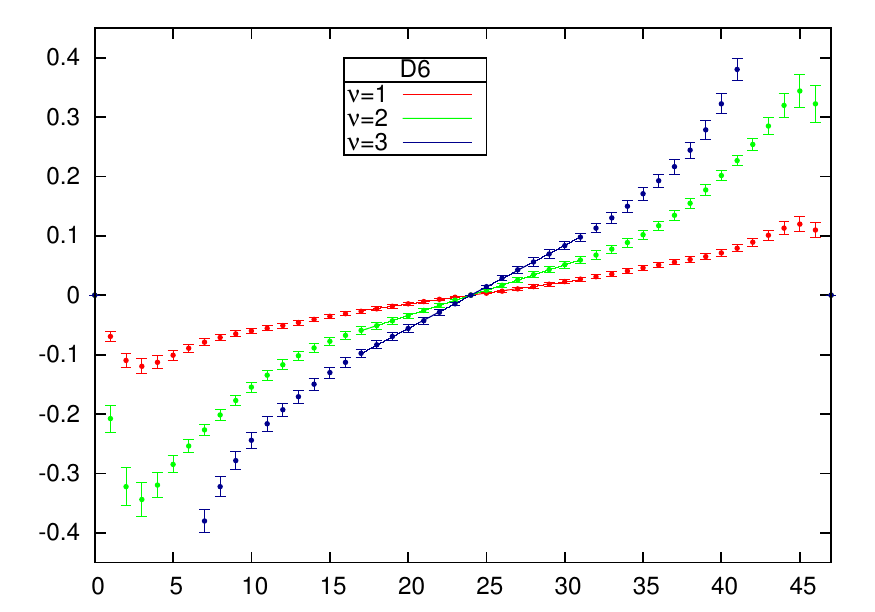}
   \end{minipage}
\caption{Plots of $D_\nu / |\nu|$ vs $t$ for two of our lattices (left : D4 -- right : D6) for several topological sectors.}
\label{fig:PPderivfig}
\end{figure}

Fig.~(\ref{fig:PPderivfig}) shows the coefficient $D_\nu / |\nu|$ as a function of time for two of our lattices and several topological charges.
This temporal dependence is very well described by the expression given in eq.~(\ref{eq:tdep}). We give some results in table~(\ref{table:Dres}) for the case of a linear fit of the $z$ dependence of $D_\nu$. We checked the stability of the fits by varying the fitting time interval and also by including higher order corrections (with a term in $z^3$). In any case, the results we obtain are in full agreement. 

\begin{figure}[h!]
   \centering \includegraphics[scale=0.8]{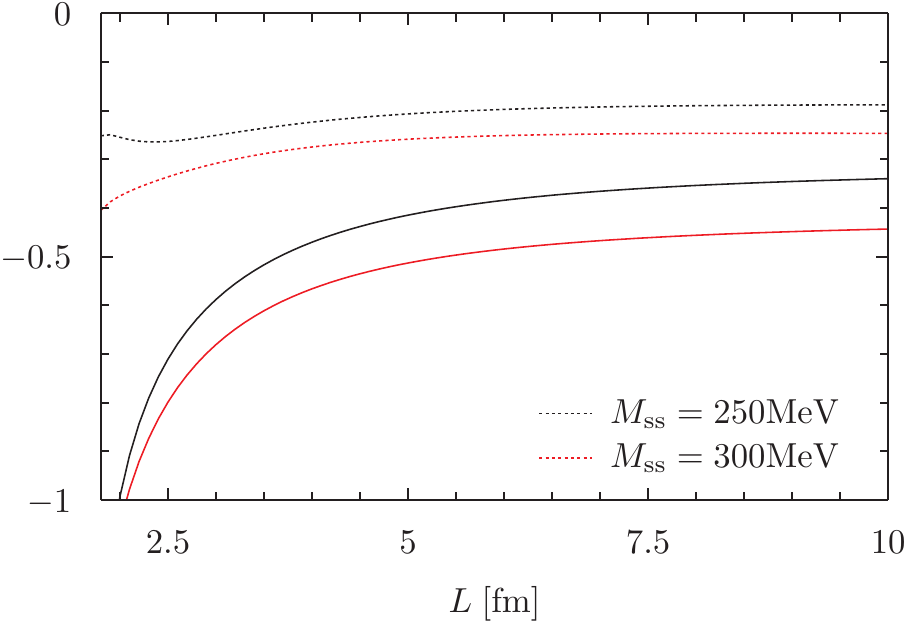}
   \caption{$(D_\nu^{NLO}/|\nu| - D_\nu^{LO}/|\nu|)/(D_\nu^{LO}/|\nu|)$ vs. $L$. Black and red lines are the results for $M_{\rm{ss}} = 250$ and $300 \; \textrm{MeV}$ while  dashed and solid curves correspond respectively to $\rho=1$ and $2$.}
\label{fig:chiralfit}
\end{figure}

\vspace*{-2mm}
Unfortunately,  the convergence of the chiral expansion is bad at this value of $L$:  for $L = 2 \; \textrm{ fm}$ and $\rho=2$, for $M_{\rm{ss}} \approx 250 \; \textrm{ MeV}$, the NLO corrections represent $100 \%$ of the LO values as illustrated by fig.~(\ref{fig:chiralfit}). It is therefore not surprising that  fits of $D_\nu$ to the NNLO prediction are not good. We expect a significantly better convergence for $\rho=1$ and $L \geq \; 3\textrm{ fm}$, with the sea masses we are considering. 
 
\section{Conclusions and outlook}

We have studied the pseudoscalar correlator in a mixed regime where the volume is large compared to the Compton wave-length of sea-quark pions but short compared to that of the valence-quark pions. ChPT can be adapted 
to this regime, and we have shown the result of a NNLO computation of this observable. In particular we have focused on the coefficient
of the $1/m_\val ^2$ contributions that appear when considering fixed-topological sectors. At the fundamental level these terms can only arise 
from the topological zero-mode contributions to the correlator. The numerical computation of these quantities relies therefore on the evaluation of the zero-mode wave functions, the full propagator is not needed. 

We have evaluated for the first time these quantities in a mixed action simulation with overlap valence quarks on $N_f=2$ improved Wilson sea fermions, with a fine-grained  but relatively small $L=1.6~\textrm{fm}$ lattice. 
We have found that these observables can be extracted robustly and efficiently. However a meaningfull matching to ChPT requires significantly  
larger volumes $L \geq 3-4 \; \textrm{ fm}$ to guarranty that the neglected higher order corrections are small.  

The mixed regime considered here can be useful to extract low-energy couplings of ChPT in a different kinematical regime with 
different systematics and different weight of the different low-energy couplings (e.g. those that give contributions proportional to the valence
quark mass would be suppressed). It might also be useful to simplify other notoriously difficult problems, such as clarifying the role of the 
charm quark mass in the $\Delta I = 1/2$ rule and extracting the weak low-energy constants relevant in $K \to \pi \pi$ decay.

\end{document}